# Design of Linear-to-Circular Polarization Transformers Made of Long Densely Packed Metallic Helices

Mário G. Silveirinha

Universidade de Coimbra, Instituto de Telecomunicações
Departamento de Engenharia Electrotécnica, Pólo II, 3030 Coimbra, Portugal, mario.silveirinha@co.it.pt

## Abstract

In this work, we study the realization of realistic polarization transformers formed by long metallic helices. To this end, we propose a new homogenization model to characterize the propagation of electromagnetic waves in medium formed by infinitely long helices. We derive approximate analytical expressions for the effective permittivity, effective permeability and the magnetoelectric tensor of the composite material, taking into account the effects of spatial dispersion. We apply the new homogenization model to characterize novel linear-to-circular polarization transformers. Our results show that the metamaterial screen may be designed in such a way that an incoming linearly polarized wave may be transformed into a circularly polarized transmitted wave, and that the transmission efficiency of such polarization transformer may be as high as 95%. The theoretical results are supported by full wave simulations.

**Index Terms:** artificial materials, polarization transformers, chiral media, helix

# I. INTRODUCTION

The characterization and study of artificial materials is an important topic in modern electromagnetics. It has been suggested that properly designed microstructured composite materials may enable the realization of compact resonators, cavities, and waveguides that may surpass diffraction limits [1, 2, 3], and allow for the selective interaction of electromagnetic waves with objects or particles much smaller than the wavelength. Other interesting potentials of these microstructured materials include the realization of imaging devices with super-resolution [4]-[8], or squeezing electromagnetic waves through subwavelength narrow channels [9] with a gigantic electric field enhancement.

The objective of this work is to characterize the electromagnetic properties of an array of infinitely long helices, and investigate its possible application to the design of polarization transformation screens. Even though the study of artificial chiral media made of metallic helices is an old subject dating back to the beginning of the 20[th] century [10]-[11], recently there has been a renewed interest in the properties of these composite materials after the suggestion of Tretyakov [12] that a racemic mixture of canonical helices may yield an effective medium with negative parameters. A related proposal/opportunity to obtain negative refraction using chiral materials was also identified in [13]. In this work we explore a different application of artificial chiral media, namely the possibility of designing a metamaterial screen to transform an incoming linear polarized wave into a circularly polarized wave.

It is well-known [11, 14, 15], that bi-isotropic chiral media have the ability to rotate the polarization state of an electromagnetic wave (gyrotropy), but that the ellipticity of the electric field (axis ratio of the polarization ellipse) remains unchanged. However, for certain applications at microwave and millimeter wave frequencies it may be interesting to transform the ellipticity of the incoming wave, rather than just to rotate its polarization state. Here, we propose a novel realistic metamaterial design of such polarization transformer. From a conceptual point of view, our approach is closely related with the proposal of [11, pp. 286], where it is suggested that a uniaxial chiral material slab may behave as a polarization transformer. Similarly, we suggest using an array of very long densely packed metallic helices as a polarization transformer, exploiting the extreme anisotropy of such metamaterial and also the effects of the chirality. The main idea, as it will be explained ahead in details, is to tailor the electromagnetic modes of the metamaterial so that one of the modes is near cut-off (due to the anisotropy of the material) and the other mode propagates with circular polarization (due to the chiral effects).

In order to characterize the interaction of electromagnetic waves with the composite material, we apply the homogenization method proposed in [16] to derive an approximate analytical model for the dielectric function of a medium formed by infinitely long metallic helices, taking into account spatial dispersion effects. We prove that under some conditions the metamaterial can be described in terms of effective permittivity and permeability tensors and a magnetoelectric tensor which is a manifestation of the chiral effects. It is shown that the results predicted by our analytical model compare well with full wave band structure calculations, even when the helices are tightly packed (with high volume fractions), or when the helix pitch is small. It is important to mention that in [17] a related spiral medium was homogenized using a local field theory. The main difference between our results and the results presented in [17] is that our model describes a completely realistic physical structure, while in [17] the derived model is only approximate since the specific microstructure of the medium is partially ignored, and the spirals are modeled using an equivalent impedance. Due to this reason, the results of [17] are limited to the case in which the radius of the helices is much smaller than the lattice constant. In a different way, the solution that we derive here does not suffer from this inconvenience, and is valid even when the helices are closely packed. As discussed ahead, only in this configuration a significant magnetoelectric coupling (which is important for the polarization-transformer application) may be induced.

We use the derived homogenization model to characterize the reflection/transmission of waves by a finite slab of the metamaterial. Our study confirms that it may be possible to design the helices and tune the thickness of the metamaterial in such a way that a linearly polarized incoming wave is transformed into a wave with circular polarization. Moreover, it is shown that the efficiency of the proposed polarization transformer can be very high, with a reflection loss as small as 5% in some cases, and a good bandwidth.

This paper is organized as follows. In section II we go over the homogenization method introduced in [16], and briefly review the constitutive relations in bianisotropic media and in spatially dispersive media. Then, in section III we use the homogenization approach proposed in [16] to extract the effective parameters of the medium formed by a square lattice of infinitely long helices. In section IV, we study the properties of the plane waves supported by the artificial medium. In section V we use the new model to investigate the opportunity of designing polarization conversion screens. Finally, in section VI the conclusion is drawn.

In this work, we assume that the fields are monochromatic and that the time dependence is $e^{j\omega t}$.

## II. GEOMETRY OF THE MATERIAL AND OVERVIEW OF THE HOMOGENIZATION METHOD

In this section, we present a brief overview of the homogenization procedure introduced in [16] and we review the constitutive relations in bianisotropic and spatially dispersive media.

## A. Geometry

To begin with, we describe the geometry of the metamaterial under study (Fig. 1). It consists of a periodic array of infinitely long perfectly electrical conducting (PEC) helices. The helices are arranged into a square lattice, with lattice constant $a$. For simplicity, we assume that the helices are embedded in air. The parameterization of the axis of the helix inside the unit cell (one turn of the helix) is:

$$\mathbf{r}_0(u) = \left( R\cos u, R\sin u, p\frac{u}{2\pi} \right), \qquad -\pi < u < \pi \qquad (1)$$

where $u$ is the parameterization variable, $R$ is the radius of the helix, and $p > 0$ when the helix is right-handed (RH) and $p < 0$ when the helix is left-handed (LH). The wire radius is $r_w$. The metamaterial is obtained by the periodic repetition of the unit cell defined by the primitive vectors $\mathbf{a}_1 = a\hat{\mathbf{u}}_x$, $\mathbf{a}_2 = a\hat{\mathbf{u}}_y$, and $\mathbf{a}_3 = a_z\hat{\mathbf{u}}_z$, as illustrated in Fig. 1. Note that the helix pitch is such that $a_z = |p|$. The helix surface $\partial D$ is parameterized as explained in Appendix A.

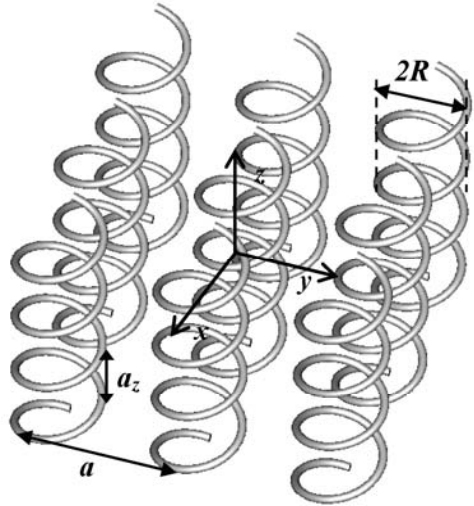

**Fig. 1** Geometry of a periodic array of infinitely long PEC helices arranged in a square lattice.

## B. Constitutive relations

Since the artificial material under study has gyrotropic properties and, as discussed ahead, may exhibit strong spatial dispersion, it is appropriate to review some elementary concepts related with the homogenization of bianisotropic media and spatially dispersive media.

The effective parameters of a periodic material are intrinsically related with the properties of its Floquet modes. As proved in [16, 18], the macroscopic average electric field $\mathbf{E}_{av}$ and the macroscopic induction field $\mathbf{B}_{av}$ associated with an electromagnetic Floquet mode $(\mathbf{E}, \mathbf{B})$ verify,

$$-\mathbf{k} \times \mathbf{E}_{av} + \omega \mathbf{B}_{av} = 0 \tag{2a}$$

$$\omega \left( \varepsilon_0 \mathbf{E}_{av} + \mathbf{P}_g \right) + \mathbf{k} \times \frac{\mathbf{B}_{av}}{\mu_0} = 0 \tag{2b}$$

where $\mathbf{k} = (k_x, k_y, k_z)$ is the wave vector characteristic of the mode, $\omega$ is the frequency, and $\mathbf{P}_g$ is the *generalized* polarization vector, which in the long wavelength limit may be written in terms of the classical polarization vector $\mathbf{P}$ (average electric dipole moment in a unit cell) and the magnetization vector $\mathbf{M}$ (averaged magnetic dipole moment in a unit cell):

$$\mathbf{P}_g = \frac{1}{V_{cell} j\omega} \int_\Omega \mathbf{J}_d e^{j\mathbf{k}\cdot\mathbf{r}} d^3\mathbf{r} \approx \mathbf{P} - \frac{\mathbf{k}}{\omega} \times \mathbf{M} \tag{3}$$

In the above, $V_{cell}$ is the volume of the unit cell, and $\mathbf{J}_d = j\omega(\varepsilon - \varepsilon_0)\mathbf{E}$ is the microscopic polarization current. The decomposition $\mathbf{P}_g \approx \mathbf{P} - \frac{\mathbf{k}}{\omega} \times \mathbf{M}$ motivates the introduction of the classical (macroscopic) electric displacement vector $\mathbf{D}_{av}$ and of the (macroscopic) magnetic field $\mathbf{H}_{av}$, through the textbook formulas:

$$\mathbf{D}_{av} = \varepsilon_0 \mathbf{E}_{av} + \mathbf{P} \tag{4a}$$

$$\mathbf{H}_{av} = \frac{\mathbf{B}_{av}}{\mu_0} - \mathbf{M} \tag{4b}$$

In particular, substituting (4) into (2b) we obtain that the classical relation between $\mathbf{D}_{av}$ and $\mathbf{H}_{av}$ for plane waves:

$$\omega \mathbf{D}_{av} + \mathbf{k} \times \mathbf{H}_{av} = 0 \tag{5}$$

For local linear (bianisotropic) media [11], it is possible to find a (relative) permittivity tensor $\overline{\overline{\varepsilon}}_r(\omega)$, a (relative) permeability tensor $\overline{\overline{\mu}}_r(\omega)$, and (dimensionless) parameters that characterize the magnetoelectric coupling $\overline{\overline{\xi}}(\omega)$ and $\overline{\overline{\zeta}}(\omega)$, such that the artificial material is characterized by the classical constitutive relations:

$$\mathbf{D}_{av} = \varepsilon_0 \overline{\overline{\varepsilon}}_r . \mathbf{E}_{av} + \sqrt{\varepsilon_0 \mu_0}\, \overline{\overline{\xi}} . \mathbf{H}_{av} \tag{6a}$$

$$\mathbf{B}_{av} = \sqrt{\varepsilon_0 \mu_0}\, \overline{\overline{\zeta}} . \mathbf{E}_{av} + \mu_0 \overline{\overline{\mu}}_r . \mathbf{H}_{av} \tag{6b}$$

The properties of the plane wave solutions in the artificial medium (dispersion relation $\omega = \omega(\mathbf{k})$ and polarization of the fields) are completely determined by (2a), (5) and by the classical constitutive relations (6).

As discussed in [16, 19], for nonlocal media (i.e. spatially dispersive media) the classical model described before may not apply, and it may be convenient to adopt different phenomenological constitutive relations. Indeed, for these materials, in general, the decomposition $\mathbf{P}_g \approx \mathbf{P} - \dfrac{\mathbf{k}}{\omega} \times \mathbf{M}$ may not either be possible or be meaningful. Thus, instead of introducing local parameters such as $\overline{\overline{\varepsilon}}_r(\omega)$ or $\overline{\overline{\mu}}_r(\omega)$, it is preferable to directly relate the average electric field $\mathbf{E}_{av}$ with the generalized polarization vector $\mathbf{P}_g$. To this end, a *generalized* displacement vector $\mathbf{D}_g$ is introduced through the relation $\mathbf{D}_g = \varepsilon_0 \mathbf{E}_{av} + \mathbf{P}_g$. The effective parameters of the composite material are then characterized by a *dielectric function*, $\overline{\overline{\varepsilon}}_{eff} = \overline{\overline{\varepsilon}}_{eff}(\omega, \mathbf{k})$, defined in such a way that:

$$\mathbf{D}_g \equiv \varepsilon_0 \mathbf{E}_{av} + \mathbf{P}_g = \overline{\overline{\varepsilon}}_{eff}(\omega, \mathbf{k}) . \mathbf{E}_{av} \tag{7}$$

Notice that the dielectric function depends on both frequency and wave vector. Substituting (7) into (2) we readily find that:

$$-\mathbf{k} \times \mathbf{E}_{av} + \omega \mathbf{B}_{av} = 0 \tag{8a}$$

$$\omega \overline{\overline{\varepsilon}}_{eff} \cdot \mathbf{E}_{av} + \mathbf{k} \times \frac{\mathbf{B}_{av}}{\mu_0} = 0 \tag{8b}$$

As discussed in [16], the plane wave solutions in the artificial medium can be completely characterized by (8) and by the dielectric function $\overline{\overline{\varepsilon}}_{eff}(\omega, \mathbf{k})$. For future reference, we note that local media characterized by the classical constitutive relations (6) can be as well described by a dielectric function of the form $\overline{\overline{\varepsilon}}_{eff} = \overline{\overline{\varepsilon}}_{eff}(\omega, \mathbf{k})$ [20]. In fact, using (6) and the decomposition $\mathbf{P}_g \approx \mathbf{P} - \frac{\mathbf{k}}{\omega} \times \mathbf{M}$ it is trivial to relate $\mathbf{P}_g$ and $\mathbf{E}_{av}$ as in (7). We obtain the following relation between the local parameters $\overline{\overline{\varepsilon}}_r(\omega)$, etc, and the dielectric function $\overline{\overline{\varepsilon}}_{eff} = \overline{\overline{\varepsilon}}_{eff}(\omega, \mathbf{k})$:

$$\frac{\overline{\overline{\varepsilon}}_{eff}}{\varepsilon_0}(\omega, \mathbf{k}) = \overline{\overline{\varepsilon}}_r - \overline{\overline{\xi}} \cdot \overline{\overline{\mu}}_r^{-1} \cdot \overline{\overline{\zeta}} + \left( \overline{\overline{\xi}} \cdot \overline{\overline{\mu}}_r^{-1} \times \frac{\mathbf{k}}{\beta} - \frac{\mathbf{k}}{\beta} \times \overline{\overline{\mu}}_r^{-1} \cdot \overline{\overline{\zeta}} \right) + \frac{\mathbf{k}}{\beta} \times \left( \overline{\overline{\mu}}_r^{-1} - \overline{\overline{\mathbf{I}}} \right) \times \frac{\mathbf{k}}{\beta} \tag{9}$$

where $\beta = \omega \sqrt{\varepsilon_0 \mu_0}$ is the free-space wave number and $\overline{\overline{\mathbf{I}}}$ is the identity dyadic. However, the converse is not true, i.e. media described by a general dielectric function $\overline{\overline{\varepsilon}}_{eff} = \overline{\overline{\varepsilon}}_{eff}(\omega, \mathbf{k})$ in general cannot be described by local parameters $\overline{\overline{\varepsilon}}_r(\omega)$, $\overline{\overline{\mu}}_r(\omega)$, etc.

In summary, both local and nonlocal (spatially dispersive) media can be always characterized by a dielectric function of the form $\overline{\overline{\varepsilon}}_{eff} = \overline{\overline{\varepsilon}}_{eff}(\omega, \mathbf{k})$, associated with the fundamental constitutive relation (7). On the other hand, local media can also be described by the classical constitutive relations (6) with local effective parameters. It is simple to verify that in unbounded media (without interfaces) both formalisms are equivalent. However, in problems involving interfaces and boundaries between different media, the classical model (6) has several important advantages as compared to the nonlocal model (7) [16, 20]. This matter will be discussed ahead in the paper.

*C. Calculation of the dielectric function*

In a recent work [16], we introduced a new, completely general and systematic approach to homogenize arbitrary non-magnetic periodic materials. The new method can fully characterize frequency dispersion, the magnetoelectric coupling (bianisotropic effects), and spatial dispersion of novel artificial media, even in frequency band-gaps or when the constituent materials are lossy. Instead of directly extracting the local parameters (permittivity, permeability, etc) of the material, the method first calculates the dielectric function $\overline{\overline{\varepsilon}}_{eff} = \overline{\overline{\varepsilon}}_{eff}(\omega, \mathbf{k})$ of the metamaterial. As discussed in [16], this approach has several advantages, namely it is computationally simpler and allows describing both local and nonlocal artificial materials in a unified manner. On the other hand, it is always possible to extract the local parameters (if they are meaningful) from the dielectric function $\overline{\overline{\varepsilon}}_{eff} = \overline{\overline{\varepsilon}}_{eff}(\omega, \mathbf{k})$. In fact, the local effective permeability and the magnetoelectric tensors may be related with the first and second order derivatives of the dielectric function at $\mathbf{k} = 0$, respectively [16, 19], as is apparent from (9).

The main idea of the method proposed in [16] is to calculate, for fixed $(\omega, \mathbf{k})$, the microscopic fields induced in the electromagnetic crystal when it is excited by a periodic external uniform harmonic electric current density of the form $\mathbf{J}_{ext} = \mathbf{J}_{e,av} e^{-j\mathbf{k}\cdot\mathbf{r}}$, where $\mathbf{k}$ is the wave vector that defines the phase shift from cell to cell, and $\mathbf{J}_{e,av}$ is a constant vector that defines the amplitude of the source (and indirectly the average macroscopic field $\mathbf{E}_{av}$). Using the calculated microscopic fields, we can then characterize the generalized polarization vector $\mathbf{P}_g$, given by (3), and afterwards it is possible to determine the dielectric function defined as in (7). For further details about the homogenization procedure the reader is referred to [16]. An important outcome of the theory presented in [16] – resulting from a method of moments (MoM) based solution of the problem – is that the dielectric function can be written is closed analytical form in terms of the expansion functions $\mathbf{w}_{n,\mathbf{k}}$ (of the induced microscopic currents) adopted in the MoM implementation. For the particular case in which the inclusions are metallic (PEC) it is proved in [16] that the dielectric function is given by:

$$\frac{\overline{\overline{\varepsilon_{eff}}}}{\varepsilon_0}(\omega,\mathbf{k}) = \overline{\overline{\mathbf{I}}} + \frac{1}{V_{cell}} \sum_{m,n} \chi^{m,n} \int_{\partial D} \mathbf{w}_{m,\mathbf{k}}(\mathbf{r}) e^{+j\mathbf{k}\cdot\mathbf{r}} ds \int_{\partial D} \mathbf{w}_{n,-\mathbf{k}}(\mathbf{r}) e^{-j\mathbf{k}\cdot\mathbf{r}} ds \qquad (10a)$$

$$\chi_{m,n} = \int_{\partial D}\int_{\partial D} \left( \nabla_s \cdot \mathbf{w}_{m,-\mathbf{k}}(\mathbf{r}) \nabla'_s \cdot \mathbf{w}_{n,\mathbf{k}}(\mathbf{r}') - \beta^2 \mathbf{w}_{m,-\mathbf{k}}(\mathbf{r}) \cdot \mathbf{w}_{n,\mathbf{k}}(\mathbf{r}') \right) \Phi_{p0}(\mathbf{r}|\mathbf{r}') ds\, ds' \qquad (10b)$$

In above $\partial D$ is the surface of the metallic inclusion in the unit cell, $\mathbf{w}_{n,\mathbf{k}}$ ($n=1,2,3…$) form a complete set of tangential vector fields defined over $\partial D$ (used for the expansion of the electric current density over the PEC surface), $\nabla_s \cdot$ stands for the surface divergence of a tangential vector field, $\beta = \omega\sqrt{\varepsilon_0 \mu_0}$ is the free-space wave number, $\Phi_{p0}$ is a Green function, and the matrix $[\chi^{m,n}]$ is the inverse of $[\chi_{m,n}]$.

The Green function $\Phi_{p0}$ may be evaluated as explained in [16] and depends on both $\omega$ and $\mathbf{k}$. In this work, we use assume that the long-wavelength limit approximation holds, i.e. that $\beta a \ll \pi$ and $|\mathbf{k}|a \ll \pi$. Using this hypothesis, and for the geometry of Fig. 1, we can write that (see [16]):

$$\Phi_{p0}(\mathbf{r}|\mathbf{r}') \approx \frac{e^{-j\mathbf{k}\cdot(\mathbf{r}-\mathbf{r}')}}{V_{cell}} \sum_{\mathbf{J}\neq 0} \frac{e^{-j\mathbf{k}_\mathbf{J}^0 \cdot (\mathbf{r}-\mathbf{r}')}}{\mathbf{k}_\mathbf{J}^0 \cdot \mathbf{k}_\mathbf{J}^0} \quad , \quad \mathbf{k}_\mathbf{J}^0 = 2\pi\left(\frac{j_1}{a}, \frac{j_2}{a}, \frac{j_3}{a_z}\right) \qquad (11)$$

where $\mathbf{J} = (j_1, j_2, j_3)$ is a generic multi-index of integers, and $\mathbf{J} \neq 0$. Notice that when written as above, $\Phi_{p0} e^{+j\mathbf{k}\cdot(\mathbf{r}-\mathbf{r}')}$ becomes independent of both frequency and wave vector. The series (11) can be efficiently summed using the approach presented in [21].

In section III we will use (10) to compute the dielectric function of the microstructured material depicted in Fig. 1. However, instead of presenting a purely numerical answer, we will derive an approximate analytical solution. This has the advantage of giving important insights into the physics of the problem and into the application of the homogenization approach.

### III. HOMOGENIZATION OF THE ARTIFICIAL MEDIUM

Next, we use the formalism presented in the previous section to homogenize the artificial material formed by an infinite square array of metallic helices (Fig. 1). Our strategy is to calculate first the dielectric function of the material (sections III.A – III.D), and afterwards to extract the effective permittivity, permeability and the magnetoelectric tensors from $\overline{\overline{\varepsilon}}_{eff}(\omega,\mathbf{k})$ (section III.E).

#### A. The expansion functions

The first step to calculate the dielectric function $\overline{\overline{\varepsilon}}_{eff}(\omega,\mathbf{k})$ given by (10) is to choose the expansion functions $\mathbf{w}_{n,\mathbf{k}}$. We remember that the current density $\mathbf{J}_c$ induced on the helix surface by the external source is written in terms of the expansion functions $\mathbf{w}_{n,\mathbf{k}}$ [16]. For simplicity, we will assume that the radius $r_w$ of the metallic wires is relatively small: $r_w \ll a$ and $r_w \ll \lambda_0$ (thin-wire approximation). Within this hypothesis, it is possible to consider that to a first approximation $\mathbf{J}_c$ flows along the helix axis [18]. Hence, the expansion functions of $\mathbf{J}_c$ may be taken equal to:

$$\mathbf{w}_{n,\mathbf{k}}(\mathbf{r}) = \frac{1}{2\pi r_w} w_n(u) \mathbf{T}(u) e^{-j\mathbf{k}\cdot\mathbf{r}} \qquad (12)$$

where $u$ is the variable associated with the parameterization $\mathbf{r}_0$ of the helix axis given by (1), $\mathbf{T} = \mathbf{r}'_0/|\mathbf{r}'_0|$ is the unit vector tangent to the helix axis, $w_n(u)$ is a generic scalar function independent of $\mathbf{k}$ to be chosen ahead, and $\mathbf{r} = \mathbf{r}(u,\phi)$ is the parameterization of the helix surface given by (A1). As explained in [16], the expansion functions $\mathbf{w}_n = \mathbf{w}_{n,\mathbf{k}}(\mathbf{r})$ must have the Floquet property. Thus (12) implies that the scalar expansion functions $w_n(u)$ are periodic functions of $u$ with fundamental period $2\pi$. Hence, a suitable set of expansion functions is the set of Fourier harmonics: $w_n(u) = \exp(jnu)$, $n = 0, \pm 1, \ldots$. As referred before, our objective here is to obtain an approximate analytical solution of the homogenization problem, rather than to obtain a numerical solution. For low frequencies it is expected that the induced current density $\mathbf{J}_c$ varies relatively slowly in one cell (see [18] for a related discussion). Hence, we expect that the behavior of the low frequency electromagnetic modes may be well described by only a few slowly varying expansion functions. Based on such physical considerations, in this work we will retain only (a linear combination of) the first 3 expansion functions. More specifically, we take the expansion functions equal to:

$$w_1(u) = -\sin u \quad ; \quad w_2(u) = \cos u \quad ; \quad w_3(u) = 1 \quad (13)$$

In what follows, we will calculate the dielectric function of the composite medium using these simplifying hypotheses. Ahead, we will compare this approximate theory with full wave simulations confirming that our simplified model yields indeed accurate results.

For future reference, we note that when the expansion functions are of the form (12) the dielectric function given by (10) can be rewritten as:

$$\frac{\overline{\overline{\varepsilon_{eff}}}}{\varepsilon_h}(\omega, \mathbf{k}) = \overline{\overline{\mathbf{I}}} + \frac{1}{V_{cell}} \sum_{m,n} \chi^{m,n} \int w_m(u) \frac{d\mathbf{r}_0}{du} du \int w_n(u') \frac{d\mathbf{r}_0}{du'} du' \quad (14)$$

where $V_{cell} = a^2 |p|$ is the volume of the unity cell. Similarly, using the results of Appendix A and (10), one finds that,

$$\chi_{m,n} = \iint du\, du' \left( \left( \frac{dw_m}{du} + j\mathbf{k} \cdot \frac{d\mathbf{r}_0}{du} w_m \right) \left( \frac{dw_n}{du'} - j\mathbf{k} \cdot \frac{d\mathbf{r}_0}{du'} w_n \right) - \beta^2 w_m(u) w_n(u') \frac{d\mathbf{r}_0}{du} \cdot \frac{d\mathbf{r}_0}{du'} \right) K_{p0}(u|u')$$

$$(15)$$

$$K_{p0}(u|u') = \frac{1}{(2\pi)^2} \int_0^{2\pi}\int_0^{2\pi} d\phi\, d\phi'\, \Phi_{p0}(\mathbf{r}|\mathbf{r}')e^{j\mathbf{k}\cdot(\mathbf{r}-\mathbf{r}')} \qquad (16)$$

Notice that in (16) we put $\mathbf{r} = \mathbf{r}(u,\phi)$ and $\mathbf{r}' = \mathbf{r}(u',\phi')$, which are defined as in (A1). We also refer that because of (11), the periodic thin-wire kernel defined above, $K_{p0}(u|u')$, is independent of both frequency and wave vector.

## B. Calculation of $\chi_{m,n}$

In order to obtain the effective permittivity using (14), next we calculate the elements $\chi_{m,n}$. To begin with, we note that $\chi_{m,n}(\mathbf{k}) = \chi_{n,m}(-\mathbf{k})$, and thus it is sufficient to calculate $\chi_{m,n}$ for $n \geq m$. These elements can be written as linear combinations of integrals of the type:

$$\int_{-\pi}^{\pi}\int_{-\pi}^{\pi} du\, du'\, K_{p0}(u|u') \times \begin{cases}\sin(mu)\\ \cos(mu)\end{cases} \times \begin{cases}\sin(nu')\\ \cos(nu')\end{cases}, \qquad m,n = 0,1,2 \qquad (17)$$

All the above integrals are identically zero, except if the function associated with the primed and unprimed coordinates is the same. This property is proved in Appendix C. For convenience, we introduce the following parameters (which only depend on the geometry of the metamaterial):

$$C_m = \int_{-\pi}^{\pi}\int_{-\pi}^{\pi} du\, du'\, \cos(mu)\cos(mu') K_{p0}(u|u'), \qquad m = 0,1,2 \qquad (18a)$$

$$\tilde{C}_m = \int_{-\pi}^{\pi}\int_{-\pi}^{\pi} du\, du'\, \sin(mu)\sin(mu') K_{p0}(u|u'), \qquad m = 1,2 \qquad (18b)$$

The unities of $C_m$ and $\tilde{C}_m$ are $[m^{-1}]$. We prove in Appendix C that $C_1 = \tilde{C}_1$. Using the previous definitions and results, we obtain after straightforward calculations that (the formulas are exact):

$$\chi_{11} = C_1 + \left(\frac{R}{2}\right)^2 (C_0 + C_2)k_x^2 + \left(\frac{R}{2}\right)^2 \tilde{C}_2 k_y^2 + \left(\frac{p}{2\pi}\right)^2 C_1 k_z^2 \\ - \beta^2\left(\left(\frac{p}{2\pi}\right)^2 C_1 + \left(\frac{R}{2}\right)^2 (C_0 + C_2 + \tilde{C}_2)\right) \qquad (19a)$$

$$\chi_{12} = j\frac{p}{2\pi}2C_1k_z + \left(\frac{R}{2}\right)^2 \left(C_0 + \tilde{C}_2 - C_2\right)k_xk_y \tag{19b}$$

$$\chi_{13} = jRC_1k_y + R\frac{p}{2\pi}\left(C_1 + \frac{C_0}{2}\right)k_xk_z \tag{19c}$$

$$\chi_{23} = -jRC_1k_x + R\frac{p}{2\pi}\left(C_1 + \frac{C_0}{2}\right)k_yk_z \tag{19d}$$

$$\chi_{33} = R^2C_1\left(k_x^2 + k_y^2\right) + \left(\frac{p}{2\pi}\right)^2 C_0k_z^2 - \left[2R^2C_1 + \left(\frac{p}{2\pi}\right)^2 C_0\right]\beta^2 \tag{19e}$$

Finally, $\chi_{22}$ is obtained from $\chi_{11}$ by interchanging the symbols $k_x$ and $k_y$, and the remaining elements can be obtained using the relation $\chi_{m,n}(\mathbf{k}) = \chi_{n,m}(-\mathbf{k})$.

## C. Calculation of $\chi^{m,n}$

Here, we calculate $\left[\chi^{m,n}\right]$, which is the inverse of the matrix $\left[\chi_{m,n}\right]$. These parameters can be written in closed analytical form, but the formulas are cumbersome. To circumvent this inconvenience, we will use the long wavelength limit approximation to simplify the formulas.

In order to clarify our approach, let us consider for example the *exact* formula for $\chi^{11}$:

$$\chi^{11} = \frac{\chi_{22}\chi_{33} - \chi_{23}\chi_{32}}{\det(\chi)} \tag{20}$$

Both the numerator and denominator of the above formula are polynomial functions of $\beta$, $k_x$, $k_y$ and $k_z$. In our calculations we only retained the lowest order terms in both the numerator and denominator. Depending on the specific $\chi^{m,n}$ these terms are monomials of order one (sum of parcels involving $\beta$, $k_x$, etc) or of order two (sum of parcels involving $\beta^2$, $k_xk_y$, $\beta k_x$, $k_x^2$, etc). Within these approximations, we find:

$$\det(\chi) \approx C_1^2 \left[\left(\frac{p}{2\pi}\right)^2 C_0k_z^2 - \left[2R^2C_1 + \left(\frac{p}{2\pi}\right)^2 C_0\right]\beta^2\right] \tag{21a}$$

$$\chi^{11} \approx \frac{1}{C_1} + \frac{C_1^2R^2k_y^2}{\det(\chi)} \tag{21b}$$

$$\chi^{12} \approx -\frac{C_1^2R^2k_xk_y}{\det(\chi)} \tag{21c}$$

$$\chi^{13} \approx \frac{-jC_1^2 R k_y}{\det(\chi)} \tag{21d}$$

$$\chi^{23} \approx \frac{+jC_1^2 R k_x}{\det(\chi)} \tag{21e}$$

$$\chi^{33} \approx \frac{C_1^2}{\det(\chi)} \tag{21f}$$

As before, $\chi^{22}$ is obtained from $\chi^{11}$ by interchanging the symbols $k_x$ and $k_y$, and the remaining elements can be obtained using the relation $\chi^{m,n}(\mathbf{k}) = \chi^{n,m}(-\mathbf{k})$.

### D. The dielectric function

To calculate $\overline{\overline{\varepsilon_{\mathit{eff}}}}$, we note that $\int w_n(u') \frac{d\mathbf{r}_0}{du'} du'$ is equal to $\pi R \hat{\mathbf{u}}_x$, $\pi R \hat{\mathbf{u}}_y$, and $p \hat{\mathbf{u}}_z$, for $n=1,2,3$, respectively. Hence, using (14) and (21) we obtain the desired dielectric function,

$$\frac{\overline{\overline{\varepsilon_{\mathit{eff}}}}}{\varepsilon_0}(\omega,\mathbf{k}) = \begin{pmatrix} \varepsilon_t - \dfrac{A^2 k_y^2}{\beta^2/\beta_{p1}^2 - k_z^2/\beta_{p2}^2} & \dfrac{A^2 k_x k_y}{\beta^2/\beta_{p1}^2 - k_z^2/\beta_{p2}^2} & \dfrac{jAk_y}{\beta^2/\beta_{p1}^2 - k_z^2/\beta_{p2}^2} \\ \dfrac{A^2 k_x k_y}{\beta^2/\beta_{p1}^2 - k_z^2/\beta_{p2}^2} & \varepsilon_t - \dfrac{A^2 k_x^2}{\beta^2/\beta_{p1}^2 - k_z^2/\beta_{p2}^2} & \dfrac{-jAk_x}{\beta^2/\beta_{p1}^2 - k_z^2/\beta_{p2}^2} \\ \dfrac{-jAk_y}{\beta^2/\beta_{p1}^2 - k_z^2/\beta_{p2}^2} & \dfrac{jAk_x}{\beta^2/\beta_{p1}^2 - k_z^2/\beta_{p2}^2} & 1 - \dfrac{1}{\beta^2/\beta_{p1}^2 - k_z^2/\beta_{p2}^2} \end{pmatrix} \tag{22}$$

where $A = \pi R^2 / p$,

$$\varepsilon_t = 1 + \frac{(\pi R)^2}{V_{\mathit{cell}}} \frac{1}{C_1} \tag{23}$$

$$\beta_{p1} = \sqrt{\frac{(2\pi p)^2}{C_0 p^2 V_{\mathit{cell}} + 8 C_1 \pi^2 R^2 V_{\mathit{cell}}}} \quad ; \quad \beta_{p2} = \sqrt{\frac{(2\pi)^2}{C_0 V_{\mathit{cell}}}} \tag{24}$$

Note that $\overline{\overline{\varepsilon_{\mathit{eff}}}}$ is a function of frequency, wave vector, radius $R$, pitch of the helix $p$, and of $A$, $\beta_{p1}$, $\beta_{p2}$ and $\varepsilon_t$, defined as above. The latter parameters are written exclusively in terms of $C_0$ and $C_1$, defined as in (18a). These two constants only depend on the geometry of the artificial material (but not on $\omega$ or $\mathbf{k}$), and are evaluated numerically. In Fig. 2 $C_0$ and $C_1$ are depicted as a function of the helix pitch for different $R$ and $r_w$. It is also worth noting that the geometrical parameter $A = \pi R^2 / p$ is positive for right-handed helices, and negative for left-handed helices. A detailed characterization of the dielectric function is presented in the next sections.

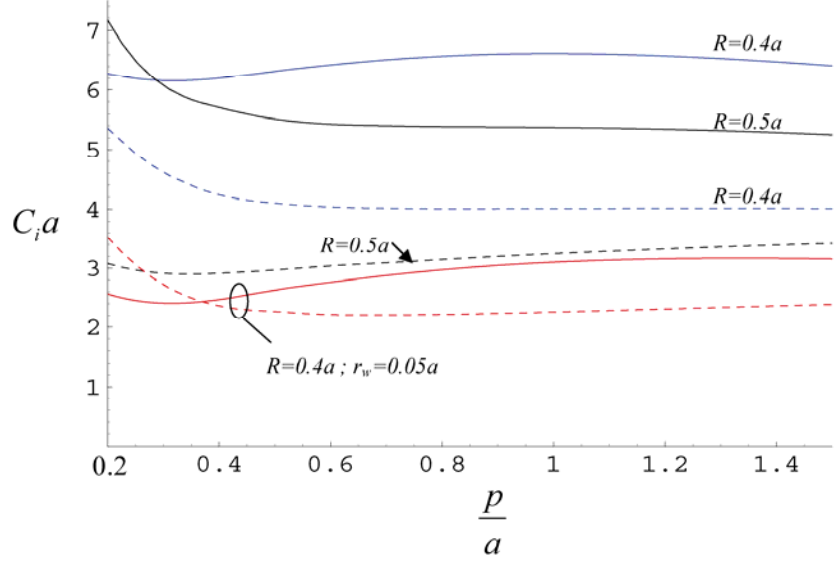

**Fig. 2** Normalized constants $C_0$ (solid lines) and $C_1$ (dashed lines) as a function of the normalized helix pitch $p$ and for different $R$. The radius of the wires is $r_w = 0.01a$ except on the curves where it is explicitly indicated that $r_w = 0.05a$.

*E. Bianisotropic model*

It is evident from (22) that the dielectric function depends explicitly on all components, $k_x$, $k_y$, and $k_z$ of the wave vector. Indeed, as explained in section II, our homogenization method is based on the constitutive relation (7), for which the effect of all microscopic currents is incorporated into the dielectric function, avoiding in this way the introduction of a permeability tensor and/or magnetoelectric coupling tensors. As discussed in [16], if the spatial dispersion is relatively weak, i.e. if the dependence of $\overline{\overline{\varepsilon}}_{eff}$ on $\mathbf{k}$ is relatively smooth so that we can expand $\overline{\overline{\varepsilon}}_{eff} = \overline{\overline{\varepsilon}}_{eff}(\omega, \mathbf{k})$ in a Taylor series in $\mathbf{k}$, it may be possible to find local parameters $\overline{\overline{\varepsilon}}_r$, $\overline{\overline{\mu}}_r$, and $\overline{\overline{\zeta}}$ such that to a first approximation (9) holds. In that case it becomes feasible to use the local constitutive relations (6) (the bianisotropic model) to characterize wave propagation in the medium. As referred in [16], the bianisotropic model – when available – is very useful because if the constitutive parameters are independent of $\mathbf{k}$, then the constitutive relations (6) are valid in the spatial domain, and consequently the classical boundary conditions (continuity of the tangential components of $\mathbf{E}$ and $\mathbf{H}$) can be used to solve boundary value problems involving interfaces with free-space [20]. Hence, the following question arises naturally: "is the model (22) equivalent to a bianisotropic model based on the constitutive relations (6)?". Or, in other words, "can we find $\overline{\overline{\varepsilon}}_r$, $\overline{\overline{\mu}}_r$, and $\overline{\overline{\zeta}}$ such that to some approximation (9) is verified?".

Unfortunately, at least if we impose that $\overline{\overline{\varepsilon}}_r$, $\overline{\overline{\mu}}_r$, and $\overline{\overline{\zeta}}$ are independent of $\mathbf{k}$, the answer to the question is negative. This can be easily demonstrated by noting that $\overline{\overline{\varepsilon}}_{eff}$ is a function of $1/\left(\beta^2/\beta_{p1}^2 - k_z^2/\beta_{p2}^2\right)$, and consequently it is obvious that it cannot be expanded into a Taylor series around $k_z = 0$ and $\omega \approx 0$. Consequently, the artificial medium under-study is characterized by strong spatial dispersion.

Even though, it is not possible to obtain $\overline{\overline{\varepsilon}}_r$, $\overline{\overline{\mu}}_r$, and $\overline{\overline{\zeta}}$ independent of $k_z$ such that to some approximation (9) holds, it is however possible to find parameters independent of $k_x$ and $k_y$ that verify (9). In fact, it can be confirmed by direct substitution that (9) is exactly verified for the following constitutive parameters:

$$\overline{\overline{\mu}}_r = \hat{\mathbf{u}}_x\hat{\mathbf{u}}_x + \hat{\mathbf{u}}_y\hat{\mathbf{u}}_y + \mu_{zz}\hat{\mathbf{u}}_z\hat{\mathbf{u}}_z, \qquad \mu_{zz} = \left(1 + \frac{\beta^2 A^2}{\beta^2/\beta_{p1}^2 - k_z^2/\beta_{p2}^2}\right)^{-1} \qquad (25a)$$

$$\overline{\overline{\zeta}} = \zeta_{zz}\hat{\mathbf{u}}_z\hat{\mathbf{u}}_z = -\overline{\overline{\xi}}^t, \qquad \mu_{zz}^{-1}\zeta_{zz} = \frac{-j\beta A}{\beta^2/\beta_{p1}^2 - k_z^2/\beta_{p2}^2} \qquad (25b)$$

$$\overline{\overline{\varepsilon}}_r = \varepsilon_t\left(\hat{\mathbf{u}}_x\hat{\mathbf{u}}_x + \hat{\mathbf{u}}_y\hat{\mathbf{u}}_y\right) + \varepsilon_{zz}\hat{\mathbf{u}}_z\hat{\mathbf{u}}_z, \qquad \varepsilon_{zz} = 1 - \frac{1}{\beta^2/\beta_{p1}^2 - k_z^2/\beta_{p2}^2} - \frac{\zeta_{zz}^2}{\mu_{zz}} \qquad (25c)$$

We refer in passing that it is possible to calculate the parameters defined above in a systematic way in terms of the derivatives of $\overline{\overline{\varepsilon}}_{eff}(\omega, \mathbf{k})$ with respect to $\mathbf{k}$. Such topic is outside the scope of the present paper. We also note that the effective parameters defined by (25) depend on the frequency, $\beta = \omega/c$, and on the wave vector component $k_z$.

Is the bianisotropic model (25) useful? As discussed in [16], the dispersion characteristic for plane waves, $\omega = \omega(\mathbf{k})$, as well as the corresponding average electric and induction fields, $\mathbf{E}_{av}$ and $\mathbf{B}_{av}$, are exactly the same independently of the considered constitutive relations, i.e. independently of using (7) and (22) or using (6) and (25). Thus, from that point of view, the bianisotropic model (25) does not bring anything new. Indeed, as noted before, the importance of the bianisotropic model derives from the fact that in general only this model can be used to solve boundary value problems [20]. However, this is true only if the associated constitutive parameters $\overline{\overline{\varepsilon}}_r$, $\overline{\overline{\mu}}_r$, and $\overline{\overline{\zeta}}$ are independent of the wave vector $\mathbf{k}$, which is not the case here since (25) depends explicitly on $k_z$.

Nevertheless, since $\overline{\overline{\varepsilon}}_r$, $\overline{\overline{\mu}}_r$, and $\overline{\overline{\zeta}}$ depend exclusively on $k_z$ (and not on $k_x$ and $k_y$), they can still be used to impose the classical boundary conditions at an interface, *provided the geometry of the problem is effectively two-dimensional (from a macroscopic point of view) and the z-direction is parallel to the interface between the metamaterial sample and free-space*, e.g., if the metamaterial sample is an arbitrarily shaped cylinder oriented along the *z*-direction and the wave propagation is confined to the *xoy* plane. In fact, in such circumstances, even though the helices are infinitely long along the *z*-direction, their footprint in the *xoy* plane is electrically small for relatively long wavelengths. Thus, for on-plane propagation the metamaterial can be homogenized using *local parameters* (i.e. non-spatially dispersive parameters). These parameters are precisely given by the proposed bianisotropic model (25). Note that since the parameters are local, i.e. independent of $k_x$ and $k_y$ for the considered 2D-macroscopic geometry, it is possible to impose the classical boundary conditions at an interface with free-space.

Hence, the importance of the bianisotropic model (25) is now obvious. Even though it is not useful to solve a generic boundary value problem in which the shape of the metamaterial block is arbitrary (as consequence of the strong spatial dispersion characteristic of the material), it can still be used to solve electromagnetic problems which are two-dimensional from a macroscopic point of view. Again, we underline that for the metamaterial under-study the model based on $\overline{\overline{\varepsilon}}_{eff} = \overline{\overline{\varepsilon}}_{eff}(\omega,\mathbf{k})$ and on the constitutive relations (7), is not useful to solve such a boundary value problem [20]. It can only be used to compute the plane wave solutions supported by an infinite periodic metamaterial. Indeed, in general spatially dispersive materials may require modified or additional boundary conditions [20, 23]. Only the bianisotropic model (25) can be used to solve a boundary value problem, as will be confirmed later in the paper.

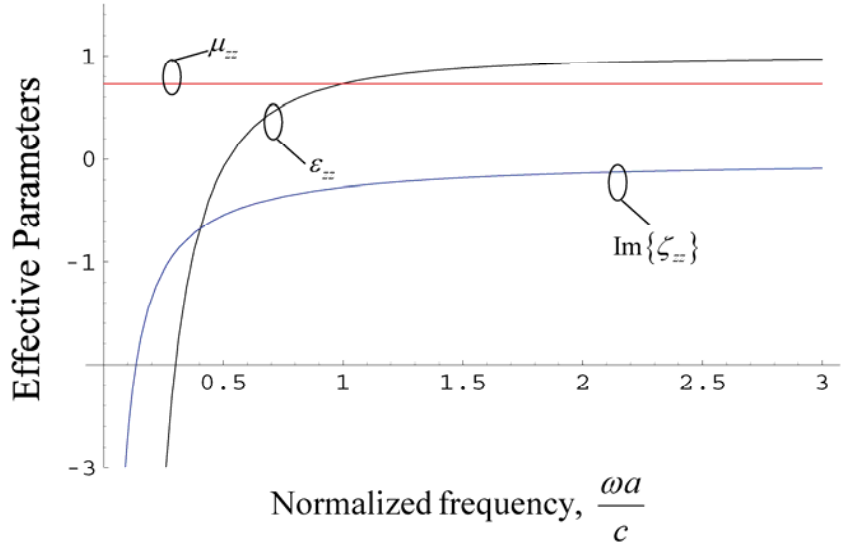

**Fig. 3** Effective parameters ($k_z = 0$) for a material with $R = 0.4a$, $r_w = 0.01a$, $p = 0.5a$.

To give an idea of the variation with frequency of the effective parameters predicted by (25), in Fig. 3 we plot $\varepsilon_{zz}$, $\mu_{zz}$ and $\text{Im}\{\zeta_{zz}\}$ (chirality), for on-plane propagation, i.e. $k_z = 0$

and for a material with $R = 0.4a$, $r_w = 0.01a$, $p = 0.5a$. It is seen that the effective permeability is less than one and is independent of frequency (within the accuracy of the proposed model and in the long wavelength limit approximation). Similarly, the transverse effective permittivity (not shown) is $\varepsilon_t = 1.77$, and is also independent of frequency. On the other hand, the effective permittivity along $z$ exhibits a plasmonic behavior. It is also seen that the chirality parameter has a resonant behavior near the static limit. For relatively low frequencies, the described results are qualitatively similar to those reported in [17]. Nevertheless, the model proposed in [17] is different from ours, possibly because in [17] the specific microstructure of the medium was partially neglected, as well as the effect of the helix polarization in the direction orthogonal to its axis.

## IV. PROPERTIES OF THE ARTIFICAL MEDIUM

The goal of this section is to characterize the plane wave solutions supported by the unbounded artificial material. Before this, it is useful to present a semi-heuristic discussion that gives some physical intuition and insights about the properties of the metamaterial under study. Thus, consider first the case in which the radius of the helices $R$ is vanishingly small, so that the artificial material reduces to the so-called "wire medium", formed by an array of parallel metallic wires oriented along the $z$-direction [22]. The wire medium supports three distinct types of electromagnetic modes: *i*) H-polarized modes (also known as TE-$z$) for which $E_z = 0$ and $H_z \neq 0$ *ii*) E-polarized modes (also known as TM-$z$) for which $E_z \neq 0$ and $H_z = 0$ *iii*) transmission line modes (also known as TEM) for which both $E_z$ and $H_z$ vanish [22]. It is well-known that for propagation in the *xoy*-plane and relatively large wavelengths the H-polarized waves propagate without interacting with the artificial medium, while the E-polarized waves are strongly attenuated.

Let us suppose that each metallic wire is slightly deformed into a helix, characterized by the radius *R* and the helix pitch *p* (see Fig. 1). How does this affect the electromagnetic modes in the metallic crystal? A helix is an object with very interesting properties. Geometrically it is a chiral object, which means that it cannot be superimposed on its mirror image [14]. It has long been known that when electromagnetic fields interact with chiral objects there may be a strong magnetoelectric coupling [11, 14], i.e. the induced electric (magnetic) dipole moment depends on both the local electric field and on the local magnetic field.

It is thus simple to understand how the winding of the metallic wires may affect the properties of the electromagnetic modes. Consider first the E-polarized mode with electric field parallel to the *z*-direction. If the straight wires are deformed into helices the current in the loop will induce a magnetic field along the *z*-direction and consequently an electric field in the *xoy* plane. So the polarization of the perturbed mode may be no longer linear but, as proved ahead, may be elliptical. The perturbed E-polarized mode is characterized by a cut-off plasma frequency.

Consider now the H-polarized mode. In this case the time varying *z*-component of the magnetic field induces an electromotive force across each turn of the winding, and so an electric field along the *z*-direction. Therefore, in general, the polarization of the perturbed mode is elliptical. An important point is that the H-polarized perturbed mode has not a cut-off frequency. In fact it propagates for arbitrarily large wavelengths, even though the electric field along the *z*-direction may be different from zero. Of course, in the static limit the polarization state is linear (i.e. the axis ratio of the polarization ellipse approaches zero as the frequency also does).

After these preliminary considerations and insights, we are ready to characterize the electromagnetic modes in the metamaterial. As referred in section III.E, the dispersion characteristic of the unbounded medium can be calculated either using the dielectric function

(22) or alternatively the bianisotropic model (25). Both approaches yield exactly the same results. We chose the first option. As proved in [16], the corresponding dispersion characteristic is given by (assuming that the electric field is not transverse, i.e. $\mathbf{k}.\mathbf{E}_{av} \neq 0$; the transverse case can be studied using a perturbation technique):

$$-1 = \mathbf{k}.\left(\beta^2 \frac{\overline{\overline{\varepsilon_{eff}}}}{\varepsilon_h} - k^2 \overline{\overline{\mathbf{I}}}\right)^{-1}.\mathbf{k} \tag{26}$$

The average macroscopic electric field associated with a solution $\beta = \beta(\mathbf{k})$ of the above equation is given by:

$$\mathbf{E}_{av} \propto \left(\frac{\overline{\overline{\varepsilon_{eff}}}}{\varepsilon_h} - \frac{k^2}{\beta^2}\overline{\overline{\mathbf{I}}}\right)^{-1}.\frac{\mathbf{k}}{\beta}, \tag{27}$$

The macroscopic induction field can be calculated using (8a).

Substituting the dielectric function (22) into (26) it can be proven that after some simplifications the characteristic equation can be rewritten in the form,

$$\beta^6 + q_2\beta^4 + q_1\beta^2 + q_0 = 0 \tag{28}$$

where $q_2$, $q_1$ and $q_0$ are polynomial functions of the wave vector $\mathbf{k}$. Since the expressions for $q_2$, $q_1$ and $q_0$ are relatively cumbersome they are not shown here (more details are given ahead for the case of on-plane propagation, which is particularly important for the design of polarization conversion screens). It is seen from (28) that the characteristic equation is a polynomial of order 3 in the variable $\beta^2$. Thus, for each wave vector $\mathbf{k}$ there are 3 different (positive) solutions for $\beta = \omega/c$, i.e. our homogenization model predicts that in the long wavelength limit 3 different modes may propagate in the metamaterial. This situation contrasts sharply with the electrodynamics of common natural media, where only two independent modes are supported (i.e. only two distinct polarizations are allowed in common natural media). The emergence of the third electromagnetic mode in the artificial material is obviously related to strong spatial dispersion. This phenomenon occurs because the helical inclusions are infinitely long objects along the *z*-direction, and thus they cannot be considered electrically small, even in the static limit. This effect is well understood and studied for the case of straight metallic wires [18], [22], [23] and was also reported in [17].

Following our initial heuristic discussion, we expect that the electromagnetic modes supported by the metamaterial are E-polarized perturbed modes, H-polarized perturbed modes, and transmission line perturbed modes. The E- and H-polarized perturbed modes were already discussed before. On the other hand, the transmission line modes correspond to waves that propagate essentially along the wires, i.e. the corresponding Poynting vector is essentially directed along the *z*-direction, independently of the transverse wave vector component $\mathbf{k}_\| = (k_x, k_y, 0)$. Note that for the case of straight wires [22], the transmission line modes are exactly dispersionless with respect to $\mathbf{k}_\|$, and propagate along the wires with the speed of light. As reported ahead, in case of helical wires the transmission line modes propagate along *z* with a velocity smaller than the speed light, since the propagation path is partially blocked by the helical turns. Also, for the case of helical wires the dispersion characteristic $\beta = \beta(\mathbf{k})$ of the transmission line modes is not exactly independent of $\mathbf{k}_\|$, even though for small frequencies such property may be valid to a first approximation. A detailed discussion of the properties of the transmission line modes is out of the scope of this paper, and will be published elsewhere.

To better understand the properties of the electromagnetic modes, it is useful to consider the particular case in which the wave travels in the *xoy*-plane (this is the situation of interest for the design of polarization transformers). For this configuration, supposing that $k_z = 0$, it is straightforward to prove that the dispersion characteristic of the 3 different families of modes (given by the solution of (26) using (22)) is,

$$\beta(k_\parallel,0) = \frac{1}{\sqrt{2\varepsilon_t}} \left[ \left((1+\varepsilon_t)k_\parallel^2 + (\varepsilon_t + A^2 k_\parallel^2)\beta_{p1}^2\right) \pm \right.$$

$$\left. \sqrt{\left(-4\varepsilon_t k_\parallel^2 (\beta_{p1}^2 + k_\parallel^2 + A^2 k_\parallel^2 \beta_{p1}^2)\right) + \left((1+\varepsilon_t)k_\parallel^2 + (\varepsilon_t + A^2 k_\parallel^2)\beta_{p1}^2\right)^2} \right]^{1/2} \quad (29a)$$

$$\beta(k_\parallel,0) = 0 \qquad \text{(transmission line modes)} \qquad (29b)$$

where $k_\parallel^2 = \mathbf{k}_\parallel \cdot \mathbf{k}_\parallel$. The two solutions associated with (29a) correspond to the E- perturbed modes ("+" sign is chosen in the equation) and to the H-polarized modes ("−" sign is chosen). From (29a) it is seen that, for on-plane propagation, the E-polarized perturbed wave is cut-off for $\beta < \beta_{p1}$. The solution (29b) is associated with the transmission line modes, and corresponds to a collapsed band such that $\beta = 0$, independently of $k_\parallel$. In fact, for very low frequencies the transmission line modes may be nearly independent of $\mathbf{k}_\parallel$ and to a good approximation have the Poynting vector along the z-direction. In these conditions, their dispersion characteristic is approximately $\beta \approx k_z/\sqrt{\varepsilon_t}$, consistently with the properties of transmission line modes in regular wire media [22].

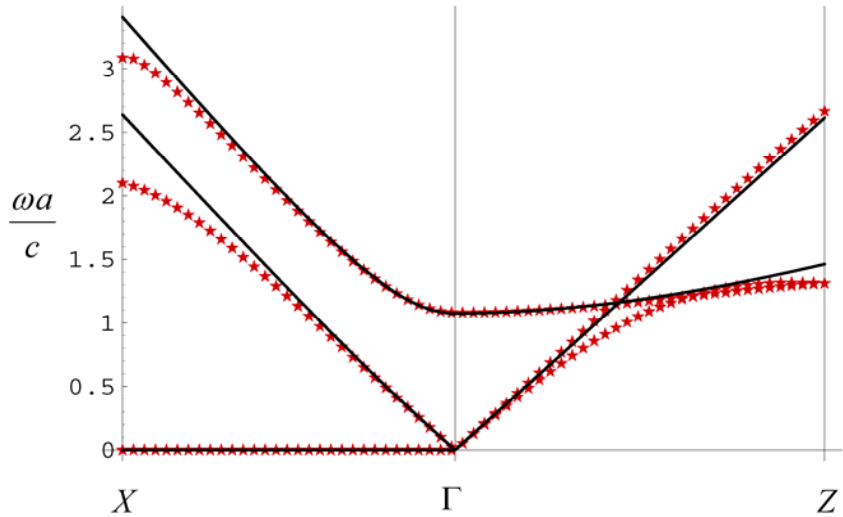

**Fig. 4** Band Structure for $R = 0.4a$, $r_w = 0.05a$, $p = 0.9a$ (only the first 3 bands are shown). Solid lines: analytical model; Star symbols: full wave numerical results. $X$, $\Gamma$ and $Z$ are highly symmetric points of the Brillouin zone.

In order to check the accuracy of the proposed analytical model and better understand the electrodynamics of the medium, we have calculated numerically the band structure of the metamaterial using the full wave hybrid-plane-wave-integral-equation method proposed in [24]. To simplify the numerical implementation of the numerical method [24], we have assumed that the thin wire approximation holds, i.e. that the induced electric current flows along the direction tangent to the helical wires. Apart from this approximation, the full wave numerical results can be considered "exact". In Fig. 4 and Fig. 5 we compare the results obtained using the analytical model (solid lines) with the full wave band structure results (star symbols). The solid lines were obtained by solving (26) using (22) (eqn. (29) is the corresponding solution for the particular case $k_z = 0$).

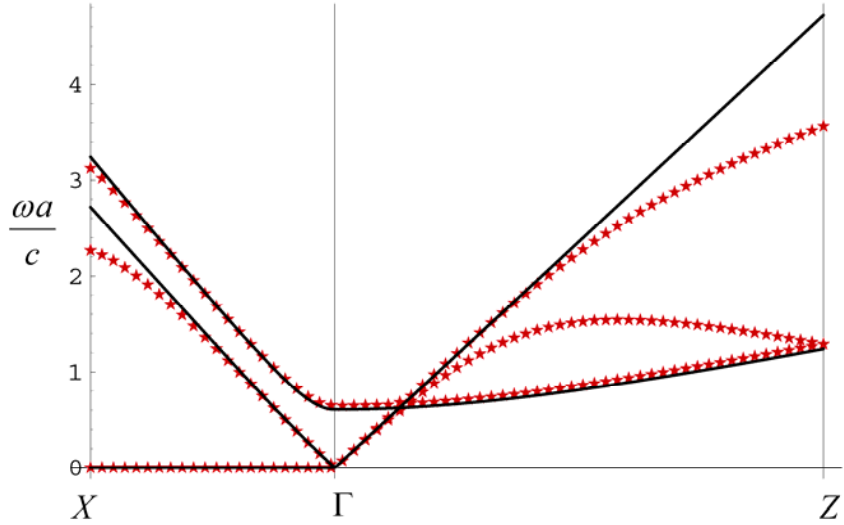

**Fig. 5** Same as in Fig. 4 but for $R = 0.4a$, $r_w = 0.01a$, $p = 0.5a$ (only the first 3 bands are shown).

In Fig. 4 and Fig. 5 the dispersion characteristic of the low-frequency electromagnetic modes is plotted along the segments $\Gamma X$ (propagation along the $x$-direction) and $\Gamma Z$ (propagation along the $z$-direction) of the Brillouin zone (BZ) (**k**-space). Note that by definition $\Gamma = (0,0,0)$, $X = \left(\dfrac{\pi}{a}, 0, 0\right)$ and $Z = \left(\dfrac{\pi}{|p|}, 0, 0\right)$. A good agreement between the theoretical model and the full wave results is revealed, especially near the $\Gamma$ point, which corresponds to propagation in the long wavelength limit.

For propagation along the $\Gamma X$ segment, with $\mathbf{k} = (k_x, 0, 0)$, the solid lines correspond to the solution given by (29). As seen it compares very well with full wave results, even for relatively large values of $k_x$ (i.e. near the $X$ point). Consistently with our initial heuristic discussion, it is seen in Fig. 4 that for low frequencies ($\beta a < \beta_{p1} a = 1.1$) the E-polarized perturbed mode is cut-off because it is strongly attenuated by the helical wires. As illustrated in Fig. 5 when the helix pitch is decreased to $p = 0.5a$ the cut-off wavenumber $\beta_{p1}$ of the E-polarized perturbed mode also decreases. Similarly, it can be verified that the cut-off frequency also decreases when the radius of the helix $R$ increases (not shown). On the other hand, Fig. 4 and Fig. 5 demonstrate that the dispersion characteristic $\beta = \beta(k)$ of the H-polarized perturbed mode is practically linear along the $\Gamma X$ segment (near the origin of Brillouin zone), and has no cut-off for low frequencies. Finally, there is a collapsed band such that $\beta = 0$, independently of $k_x$. This band corresponds to transmission line modes, as discussed before.

For propagation along the $\Gamma Z$ segment, with $\mathbf{k} = (0, 0, k_z)$, the situation is different. Indeed, the analytical model predicts that both the H-polarized and the transmission line perturbed modes are degenerate along $z$, and thus should propagate with the same phase velocity, independent of frequency (the corresponding dispersion characteristic is $\beta(0,0,k_z) \approx k_z / \sqrt{\varepsilon_t}$). This prediction is supported by the full wave results, but only for relatively low frequencies (roughly, $\beta a < 0.8$ in both figures). For higher frequencies, the full wave results demonstrate that the analytical model is not accurate since the dispersion characteristics of the two modes become clearly distinct. Moreover the modes may enter cut-off due to the emergence of electromagnetic band-gaps.

We have also characterized the polarization of the E- and H-perturbed modes, assuming propagation along the *x*-direction. Simple calculations demonstrate that (27) predicts that the polarization of the fields is proportional to:

$$\mathbf{E}_{av}(k_x) = jAk_x \frac{\beta_{p1}^2}{\beta^2}\hat{\mathbf{u}}_y + \left(\varepsilon_t - \frac{\beta_{p1}^2}{\beta^2}A^2k_x^2 - \frac{k_x^2}{\beta^2}\right)\hat{\mathbf{u}}_z, \qquad k_y = k_z = 0 \qquad (30)$$

The above formula is valid for both the E- and H-polarized perturbed modes (of course, from (29a) it is clear that for each family of modes the relation between $\beta$ and $k_x$ is different). This result confirms that provided $k_x$ is a real number (i.e. for propagating modes) the wave is transverse electromagnetic (TEM) with elliptic polarization, being the principal axes of the polarization ellipse along the *y*- and *z*-directions. Moreover, it can be easily proven that the H-perturbed mode has left-handed polarization if the helices are right-handed ($p > 0$, or equivalently $A > 0$), and has right-handed polarization if the helices are left-handed ($p < 0$). Conversely, the polarization of E-perturbed mode has the same handiness as the helices (provided it is a propagating mode with $k_x$ a real number). Thus, the E- and H-polarized perturbed modes rotate in opposite directions. Finally, we note that when the E-polarized perturbed mode is cut-off ($\beta < \beta_{p1}$) the polarization of the corresponding wave is linear.

In order to validate the previous results we have numerically calculated the polarization of the electromagnetic modes using the full wave numerical method [24]. To this end, the electromagnetic modes are obtained numerically and the electric field is then averaged over the unit cell. The comparison between the full wave results (discrete symbols) and (30) (solid lines) is depicted in Fig. 6, where we plot the axis ratio of the polarization ellipse as a function of the normalized frequency for different metamaterial configurations. It is seen that the general agreement is good, even though the helices are closely packed ($R = 0.4a$) and the helix pitch can be as small as $p = 0.2a$ (curve c). Note that we defined the axis ratio (AR) of the polarization ellipse as the ratio between the minor axis of the ellipse and the major axis of the ellipse, and so $AR \leq 1$ (in many textbooks $AR$ is defined as the inverse of this quantity).

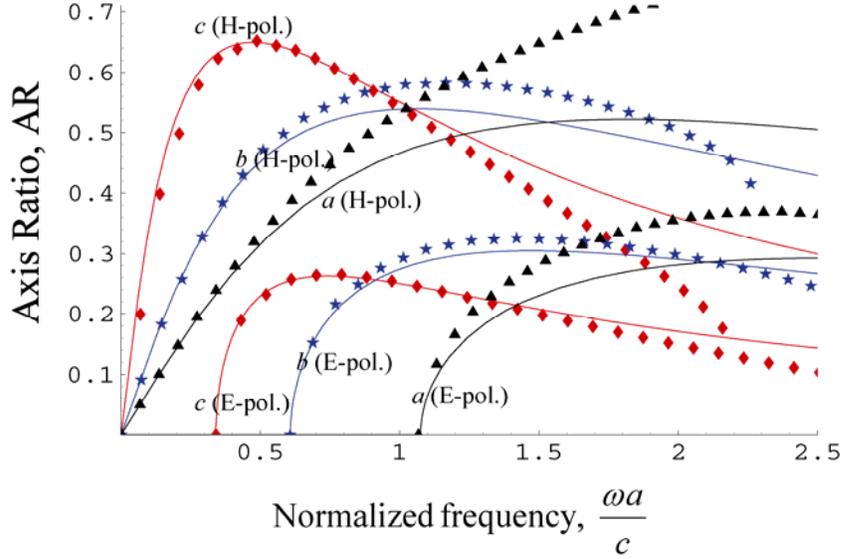

**Fig. 6** Axis ratio of the E-polarized and H-polarized electromagnetic modes as function of the normalized frequency (propagation along the *x*-direction is implicit). Solid lines: Analytical model; Diamond/Triangle/Star symbols: Full wave results. The radius of the helices is $R = 0.4a$. The wire radius and the helix pitch are: *a*) $r_w = 0.05a$, $p = 0.9a$; *b*) $r_w = 0.01a$, $p = 0.5a$; *c*) $r_w = 0.01a$, $p = 0.2a$.

From Fig. 6 it is apparent that the $AR$ for the H-polarized perturbed mode is maximal near the plasma frequency where the E-polarized perturbed mode starts propagating, $\beta \approx \beta_{p1}$; note that by decreasing the pitch $p$ the value of $\beta_{p1}$ also decreases. The results of Fig. 6 suggest that by properly designing the inclusions it may be possible to design a screen made of long metallic helices such that at some frequency the E-polarized perturbed mode is cut-off, whereas the H-polarized perturbed mode propagates with circular polarization ($AR = 1$). This may be possible if the magnetoelectric coupling is very strong, and thus if the radius of the helices $R$ is large. In such conditions, if a linearly polarized plane wave impinges on a metamaterial screen (with polarization in the *xoy* plane) it will excite an eigenmode in the metamaterial (desirably) characterized by circular polarization. This suggests the possibility of designing a linear to circular polarization conversion screen. This opportunity will be investigated next.

## V. POLARIZATION TRANSFORMERS

In what follows we study the possibility of the exploiting the magnetoelectric coupling intrinsic of the helical inclusions to design linear to circular polarization transformers. We explain how to the derived homogenization model can be used to characterize the scattering of plane waves by a finite slab of the metamaterial, and how the helices can be properly designed, and the thickness of the metamaterial screen tuned, in order that an incoming wave with linear polarization is transformed into a wave with circular polarization. To begin with, we will study with more detail the properties of the electromagnetic modes supported by the artificial medium.

### A. Propagation along the x-direction

In order to characterize the fields scattered from a slab of the artificial material formed by the metallic helices, next we study the properties of its electromagnetic modes with further detail for the case of propagation along the *x*-direction, i.e. with $k_y = k_z = 0$. For a given (normalized) frequency $\beta = \omega/c$, the E- and H-polarized perturbed modes propagate along the *x*-direction with propagation constant $k_x^E$ and $k_x^H$, respectively. These propagation constants can be calculated by solving (26) with respect to $k_x$ (or equivalently by solving (29a) with respect to $k_x$). Straightforward calculations yield:

$$k_x^{(H,E)} = \frac{1}{\sqrt{2 + 2A^2\beta_{p1}^2}} \left[ \beta^2\left(1 + A^2\beta_{p1}^2 + \varepsilon_t\right) - \beta_{p1}^2 \pm \right.$$
$$\left. \sqrt{4\beta^2\left(\beta_{p1}^2 - \beta^2\right)\left(1 + A^2\beta_{p1}^2\right)\varepsilon_t + \left(\beta^2\left(1 + A^2\beta_{p1}^2 + \varepsilon_t\right) - \beta_{p1}^2\right)^2} \right]^{1/2} \quad (31)$$

where the "+" sign is chosen is chosen for the H-perturbed modes and the "−" sign is chosen for the E-perturbed modes.

As discussed in section IV, the average electric field of the modes is given by (30). On the other hand, the corresponding average induction field verifies $\eta_0 \dfrac{\mathbf{B}_{av}}{\mu_0} = \dfrac{\mathbf{k}}{\beta} \times \mathbf{E}_{av}$ ($\eta_0 = \sqrt{\mu_0/\varepsilon_0}$ is the impedance of free-space), and so we have that,

$$\eta_0 \frac{\mathbf{B}_{av}}{\mu_0} = -\left(\varepsilon_t - \frac{\beta_{p1}^2}{\beta^2}A^2 k_x^2 - \frac{k_x^2}{\beta^2}\right)\frac{k_x}{\beta}\hat{\mathbf{u}}_y + jAk_x \frac{\beta_{p1}^2}{\beta^2}\frac{k_x}{\beta}\hat{\mathbf{u}}_z \qquad (32)$$

From (6b) the macroscopic magnetic field is given by:

$$\eta_0 \mathbf{H}_{av} = \overline{\overline{\mu}}_r^{-1} \cdot \eta_0 \frac{\mathbf{B}_{av}}{\mu_0} - \overline{\overline{\mu}}_r^{-1} \cdot \overline{\overline{\zeta}} \cdot \mathbf{E}_{av} \qquad (33)$$

Substituting (30) and (32) in the above, it follows that,

$$\eta_0 \mathbf{H}_{av}(k_x) = -\left(\varepsilon_t - \frac{\beta_{p1}^2}{\beta^2}A^2 k_x^2 - \frac{k_x^2}{\beta^2}\right)\frac{k_x}{\beta}\hat{\mathbf{u}}_y + \left(j\frac{Ak_x}{\mu_{zz}}\frac{\beta_{p1}^2}{\beta^2}\frac{k_x}{\beta} - \frac{\zeta_{zz}}{\mu_{zz}}\left(\varepsilon_t - \frac{\beta_{p1}^2}{\beta^2}A^2 k_x^2 - \frac{k_x^2}{\beta^2}\right)\right)\hat{\mathbf{u}}_z \qquad (34)$$

where $\mu_{zz}$ and $\zeta_{zz}$ are given by (25).

## B. Scattering Problem

Here we explain how the analytical model can be used to characterize the scattering of plane waves by a finite slab of the metamaterial. The slab is infinite and periodic along the y-direction (the lattice constant is *a*), and has $N_L$ layers of helices arranged along the x-direction. Thus the equivalent thickness of the slab is $L = N_L a$. The geometry for the case $N_L = 2$ is depicted in the inset of Fig. 7. The interfaces with free-space are taken equal to $x = 0$ and $x = L$. Thus, the helices are centered at $(a/2 + na, ma, 0)$, with *m* integer and $n = 0, 1, \ldots N_L - 1$.

We assume that the incident wave propagates along the *x*-direction (normal incidence) and is polarized along the *y*-direction, $\mathbf{E}^{inc} = E^{inc} e^{-jk_x^{inc} x} \hat{\mathbf{u}}_y$, where $k_x^{inc} \equiv \beta$ is the propagation constant of the incoming wave. The total electromagnetic fields for $x < 0$ are given by:

$$\mathbf{E} = E^{inc} e^{-jk_x^{inc} x} \hat{\mathbf{u}}_y + E^{inc} \left( \rho_{co} \hat{\mathbf{u}}_y + \rho_{cr} \hat{\mathbf{u}}_z \right) e^{+jk_x^{inc} x} \tag{35a}$$

$$\eta_0 \mathbf{H} = E^{inc} e^{-jk_x^{inc} x} \hat{\mathbf{u}}_z + E^{inc} \left( \rho_{cr} \hat{\mathbf{u}}_y - \rho_{co} \hat{\mathbf{u}}_z \right) e^{+jk_x^{inc} x}, \qquad x < 0 \tag{35b}$$

where $\rho_{co}$ is the reflection coefficient for the co-polarized wave, and $\rho_{cr}$ is the reflection coefficient for the cross-polarized wave. Similarly, for $x > L$, the transmitted fields are of the form,

$$\mathbf{E} = E^{inc} \left( t_{co} \hat{\mathbf{u}}_y + t_{cr} \hat{\mathbf{u}}_z \right) e^{-jk_x^{inc} x} \tag{36a}$$

$$\eta_0 \mathbf{H} = E^{inc} \left( -t_{cr} \hat{\mathbf{u}}_y + t_{co} \hat{\mathbf{u}}_z \right) e^{-jk_x^{inc} x}, \qquad x > L \tag{36b}$$

where $t_{co}$ is the transmission coefficient for the co-polarized wave, and $t_{cr}$ is the transmission coefficient for the cross-polarized wave.

On the other hand, inside the metamaterial slab, $0 < x < L$, we can write the fields in terms of E- and H-polarized perturbed modes predicted by our homogenization model. Note that the transmission line modes cannot be excited for normal incidence. Hence, we have the following expansions:

$$\mathbf{E} = c_E^+ \mathbf{E}_{av}\left(k_x^E\right) e^{-jk_x^E x} + c_E^- \mathbf{E}_{av}\left(-k_x^E\right) e^{+jk_x^E x} + c_H^+ \mathbf{E}_{av}\left(k_x^H\right) e^{-jk_x^H x} + c_H^- \mathbf{E}_{av}\left(-k_x^H\right) e^{+jk_x^H x} \tag{37a}$$

$$\mathbf{H} = c_E^+ \mathbf{H}_{av}\left(k_x^E\right) e^{-jk_x^E x} + c_E^- \mathbf{H}_{av}\left(-k_x^E\right) e^{+jk_x^E x} + c_H^+ \mathbf{H}_{av}\left(k_x^H\right) e^{-jk_x^H x} + c_H^- \mathbf{H}_{av}\left(-k_x^H\right) e^{+jk_x^H x}, \tag{37b}$$

where $\mathbf{E}_{av}(k_x)$, $k_x^{(E,H)}$, and $\mathbf{H}_{av}(k_x)$ are defined by (30), (31), and (34), respectively, and $c_{E,H}^\pm$ are the unknown amplitudes of the excited modes.

In order to determine the reflection coefficients $\rho_{co}$ and $\rho_{cr}$, the transmission coefficients $t_{co}$ and $t_{cr}$, and the four coefficients $c_{E,H}^\pm$, we need to impose that the tangential electromagnetic fields ($E_y, E_z, H_y, H_z$) are continuous at the interfaces $x = 0$ and $x = L$. In this way, we obtain an 8×8 linear system that can be solved numerically with respect to the unknowns. The results of the numerical simulations are described in section V.C.

## C. Design of polarization transformers

In what follows we prove that it may be possible to take advantage of strong magnetoelectric coupling of the artificial medium to design polarization conversion screens. As it is manifest from (37), when an incoming wave illuminates a metamaterial screen it may excite both the E- and H- polarized perturbed modes. As discussed in section IV, the H-perturbed wave has elliptical polarization, and the E-perturbed wave has either linear polarization ($\beta < \beta_{p1}$, when the mode is attenuated due to the interaction with the helices) or elliptical polarization ($\beta > \beta_{p1}$, above the plasma frequency). Our intuition is that for $\beta < \beta_{p1}$, when the E-perturbed mode is cut-off, an incoming wave (with electric field in the *xoy* plane) will couple most of its energy to the H-perturbed wave, and consequently the metamaterial screen may behave as a very efficient polarization transformer. As discussed before, it is seen in Fig. 6 that the polarization ellipse of the H-perturbed mode may have an $AR$ relatively close to unity near $\beta \approx \beta_{p1}$, which suggests that the polarization of the transmitted wave may be nearly circular.

It is important to mention that the characterization and design of polarization transformers exploiting the anisotropy and chirality of materials was studied in [11, pp. 286]. It was found that for a fixed frequency the maximal change in the ellipticity of a plane wave as it propagates in an unbounded anisotropic chiral material is achieved at a distance $L = \pi / \left( k_x^H - k_x^E \right)$ (in our notations) from a given reference plane [11, pp. 290]. The previous formula can be regarded as rough estimation of the required metamaterial thickness (of course this condition alone does not guarantee the desired linear to circular polarization conversion). It implies that to design a thin polarization transformer it is necessary that $k_x^H - k_x^E$ is as large as possible (the formula assumes that both propagation constants are real; this may not be the case in our problem). From Fig. 4 (for example) it is seen that $k_x^H - k_x^E$ is maximum at the plasma frequency $\beta = \beta_{p1}$, when $k_x^E = 0$. This suggests that to design a thin polarization transformer one should operate the metamaterial around the plasma frequency $\beta = \beta_{p1}$.

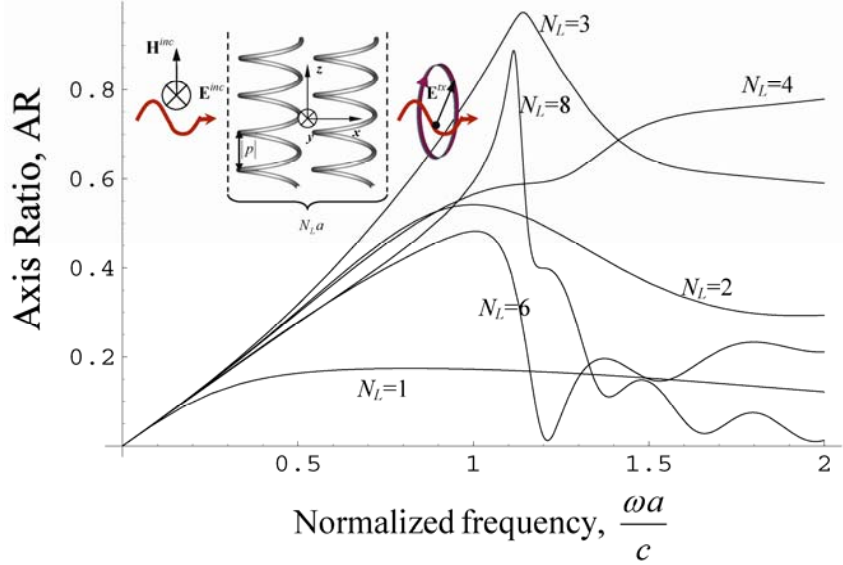

**Fig. 7** Axis ratio of the transmitted wave as function of the normalized frequency varying the number of layers $N_L$ of the metamaterial screen. The radius of the helices is $R = 0.4a$, the wire radius is $r_w = 0.05a$, and the helix pitch is $p = 0.9a$. The inset shows the geometry of the problem for the case $N_L = 2$ layers.

To have a more precise idea of the suggested possibilities we solved numerically the scattering problem formulated in section V.B, using the new homogenization model. We computed the $AR$ of the transmitted wave as a function of frequency for a screen with $R = 0.4a$, $r_w = 0.05a$, and $p = 0.9a$. The result is depicted in Fig. 7 for different number of layers of helices $N_L$. We refer that the principal axes of the polarization ellipse of the transmitted waves are not in general aligned with the Cartesian $y$ and $z$ axes. As seen in Fig. 7, the $AR$ of the transmitted wave depends considerably on the thickness of the screen. This is understandable because by varying $N_L$ we may enhance or deteriorate the $AR$ of the transmitted wave due to the interference between the different modes excited in the metamaterial slab. Consistently with our previous discussion and the results of Fig. 6, the peak of the $AR$ is attained at $\beta \approx \beta_{p1}$, which for this example corresponds to $\beta_{p1} a \approx 1.1$. It is noticeable that for $N_L = 3$ layers and $\beta \approx \beta_{p1}$ the axis ratio of the polarization ellipse is $AR \approx 1$, i.e. for these parameters the metamaterial screen behaves, in fact, as a linear to circular polarization transformer. Note that due to the interference between the modes inside the slab, the $AR$ of the transmitted wave may be much larger than the $AR$ of the H-polarized perturbed mode (Fig. 6).

In order to validate the results suggested by Fig. 7 (obtained using homogenization theory), we used the periodic method of moments (MoM) to calculate the $AR$ of the transmitted wave. The corresponding full wave numerical results (discrete symbols) are depicted in Fig. 8 and Fig. 9 superimposed on the analytical model results (solid lines), for different metamaterial geometries.

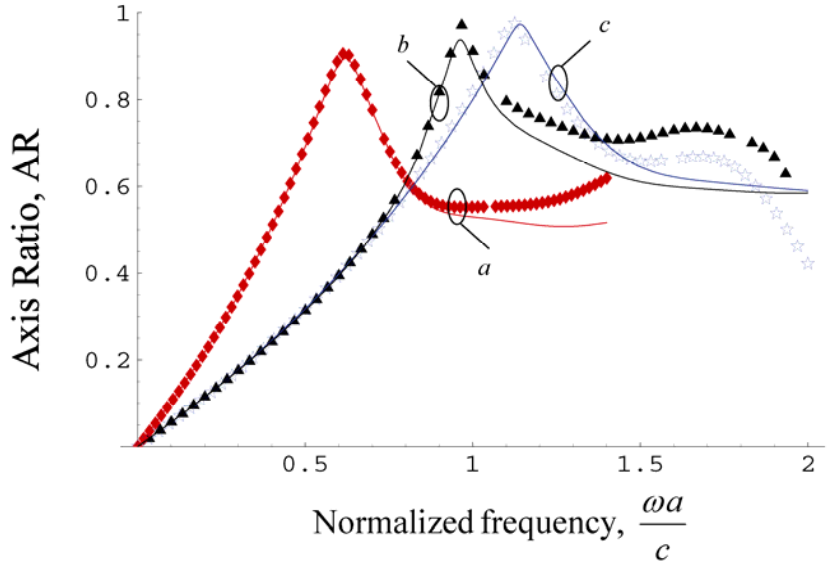

**Fig. 8** Axis ratio of the transmitted wave as function of the normalized frequency. Solid lines: Analytical model; Diamond/Triangle/Star symbols: Full wave results. The radius of the helices is $R = 0.4a$. The wire radius, helix pitch, and the number of layers are: a) $r_w = 0.01a$, $p = 0.5a$, $N_L = 5$; b) $r_w = 0.01a$, $p = 0.9a$, $N_L = 5$; c) $r_w = 0.05a$, $p = 0.9a$, $N_L = 3$.

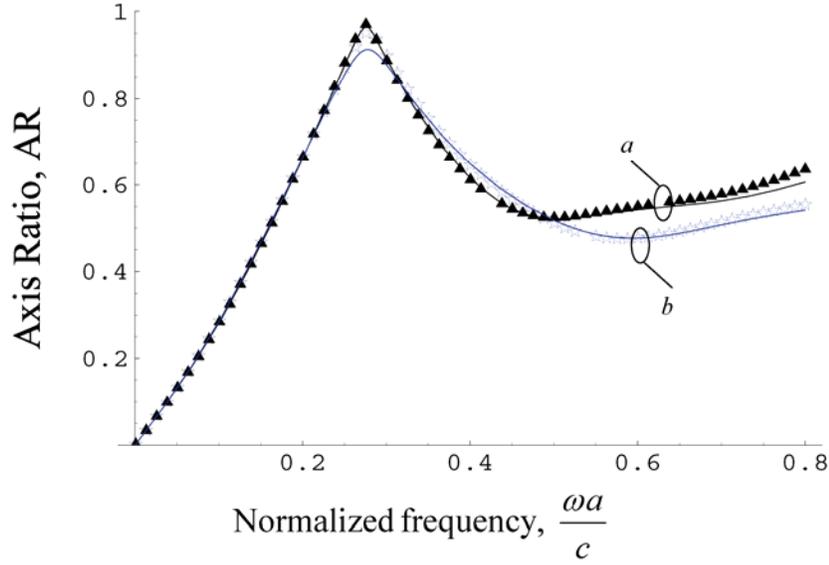

**Fig. 9** Axis ratio of the transmitted wave as function of the normalized frequency. Solid lines: Analytical model; Triangle/Star symbols: Full wave results. The radius of the helices is $R = 0.4a$ and the helix pitch is $p = 0.2a$. The wire radius, and the number of layers are: a) $r_w = 0.01a$, $N_L = 8$; b) $r_w = 0.05a$, $N_L = 6$.

The Figures reveal that despite the helices are closely packed ($R = 0.4a$) and the helix pitch is as small as $p = 0.2a$ (Fig. 9), the agreement between the analytical model and the full wave results is excellent. In all the examples plotted in Fig. 8 and Fig. 9, we have chosen the minimum value for $N_L$ that ensures that the $AR$ of the transmitted wave is close to unity at some frequency (in general $\beta \approx \beta_{p1}$). The numerical results demonstrate that the incoming linear wave can indeed be transformed into a circularly polarized wave. Notice that in the examples considered here $p > 0$ (the helices are right-handed) and, consequently, the polarization of the transmitted wave is left-handed (see the discussion in section IV). Obviously, when the helices are wounded in the opposite direction so that $p < 0$, the polarization of the transmitted wave becomes right-handed.

Fig. 8 and Fig. 9 also demonstrate that by decreasing the helix pitch $p$, the $AR = 1$ condition occurs at a smaller normalized frequency (because the normalized plasma frequency

$\beta_{p1}a$ decreases due to the enhanced magnetoelectric coupling). However, the number of layers $N_L$ necessary to reveal the effect increases.

An important parameter to characterize the efficiency of the polarization conversion is the fraction of transmitted power, given by $T = 1 - |\rho_{co}|^2 - |\rho_{cr}|^2$, where $\rho_{co}$ is the reflection coefficient for the co-polarized wave, and $\rho_{cr}$ is the reflection coefficient for the cross-polarized wave (see (35)). We calculated $T$ as a function of the normalized frequency using both the analytical model of section V.B and the periodic method of moments. The results associated with Fig. 8 and Fig. 9 are depicted in Fig. 10 and Fig. 11, and confirm the very good agreement between the homogenization results (solid lines) and the full wave results (discrete symbols). It is seen that for some of the geometries the transmission efficiency can be nearly 100% at the frequency of interest (in particular this true for curve $b$ in Fig. 10, which corresponds to $r_w = 0.01a$, $p = 0.9a$, $N_L = 5$). The results of Fig. 11 indicate that the transmission efficiency may deteriorate if the helix pitch is too small, even though a transmission as high as 70% is still possible at the frequency where $AR \approx 1$. In any case, the results presented here clearly demonstrate the potentials of the proposed metamaterial screens as polarization transformers, and prove their robustness both in terms of frequency response and transmission efficiency.

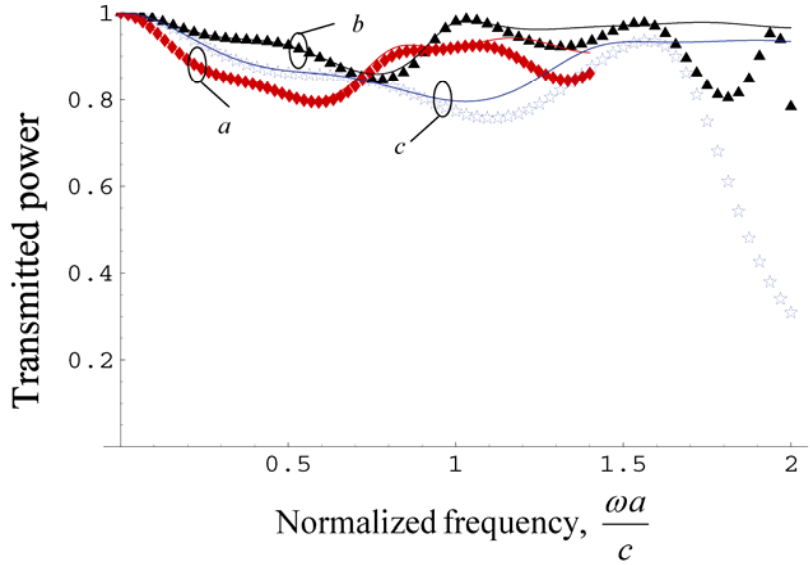

**Fig. 10** Fraction of the transmitted power as a function of the normalized frequency. Solid lines: Analytical model; Diamond/Triangle/Star symbols: Full wave results. The parameters for curves *a*, *b* and *c* are as in Fig. 8.

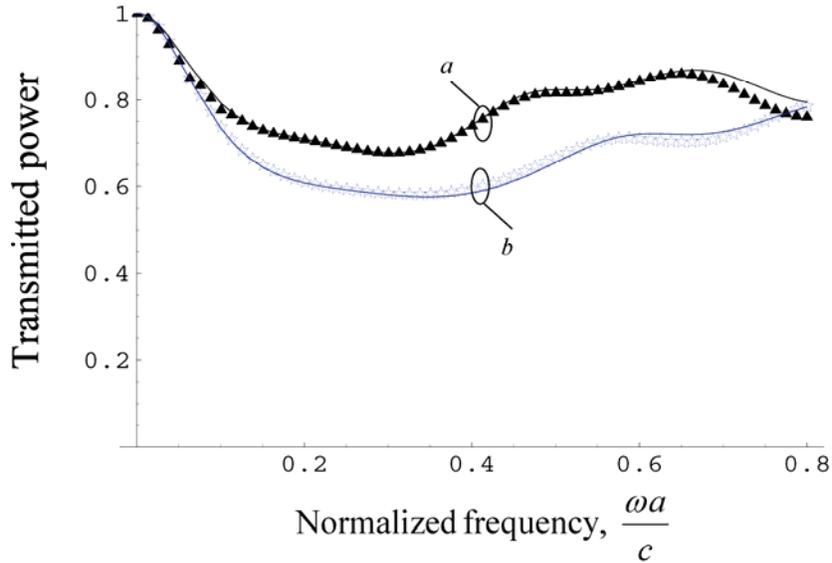

**Fig. 11** Fraction of the transmitted power as a function of the normalized frequency. Solid lines: Analytical model; Triangle/Star symbols: Full wave results. The parameters for curves *a*, and *b* are as in Fig. 9.

## VI. CONCLUSION

In this work, we studied the realization of polarization transformers using realistic microstructured materials. To this end, we developed a new homogenization model to characterize a periodic medium formed by long metallic helices. We extracted the effective permittivity, permeability and chirality of the metamaterial. We compared the proposed analytical model with full wave band structure simulations showing good agreement, even for demanding cases in which the helices are densely packed and the pitch is relatively small. We applied the results to the design of novel polarization transformers. It was shown that the proposed metamaterial may act effectively as a linear-to-circular polarization transformer, with low reflection loss, good bandwidth, and yielding a transmitted wave with axis ratio very close to one. The theoretical results are supported by full wave simulations.

## APPENDIX A

Here, we present the details about the mathematical model of the helices. The parameterization of the surface of a generic helix is:

$$\mathbf{r}(u,\phi) = \mathbf{r}_0(u) + r_w \left( \cos\phi \, \mathbf{N}(u) + \sin\phi \, \mathbf{B}(u) \right), \qquad 0 < \phi < 2\pi \qquad (A1)$$

where $r_w$ is the wire radius, $\mathbf{r}_0(u)$ is a parameterization of the helix axis given by (1), and $\mathbf{N}$ and $\mathbf{B}$ are the associated normal and binormal vectors [25]. The tangent vector is $\mathbf{T} = \mathbf{r}_0' / |\mathbf{r}_0'|$. The Frenet-Serret formulas state that [25]:

$$\begin{pmatrix} \mathbf{T}' \\ \mathbf{N}' \\ \mathbf{B}' \end{pmatrix} = |\mathbf{r}_0'| \begin{pmatrix} 0 & K & 0 \\ -K & 0 & \tau \\ 0 & -\tau & 0 \end{pmatrix} \begin{pmatrix} \mathbf{T} \\ \mathbf{N} \\ \mathbf{B} \end{pmatrix} \qquad (A2)$$

where $K$ is the curvature and $\tau$ is the torsion of the curve. For a helix, $K = R / \left( R^2 + \left( \frac{p}{2\pi} \right)^2 \right)$ and $\tau = \frac{p}{2\pi} / \left( R^2 + \left( \frac{p}{2\pi} \right)^2 \right)$. Using these formulas we readily find that:

$$\frac{\partial \mathbf{r}}{\partial u} = |\mathbf{r}_0'| \left[ \mathbf{T} + r_w \left( \cos\phi \left( -K\mathbf{T} + \tau\mathbf{B} \right) - \sin\phi \, \tau \mathbf{N} \right) \right] \qquad (A3)$$

$$\frac{\partial \mathbf{r}}{\partial \phi} = r_w \left( -\sin\phi \, \mathbf{N} + \cos\phi \, \mathbf{B} \right) \qquad (A4)$$

$$ds = \sqrt{g} \, du \, d\phi, \qquad \sqrt{g} = r_w |\mathbf{r}_0'| \left( 1 - K r_w \cos\phi \right) \qquad (A5)$$

where $ds$ is the area of a surface element. Within the thin-wire approximation the density of current $\mathbf{J}_c$ is of the form (a similar property holds for the expansion functions $\mathbf{w}_n$):

$$\mathbf{J}_c = \frac{1}{2\pi} \frac{I(u)}{\sqrt{g}} \frac{\partial \mathbf{r}}{\partial u} \tag{A6}$$

where $I(u)$ is the current along the wires. Then, we have that:

$$\nabla_S \cdot \mathbf{J}_c = \frac{1}{\sqrt{g}} \frac{1}{2\pi} \frac{dI}{du} \tag{A7}$$

In this work it is assumed that $K r_w \ll 1$ and $\tau r_w \ll 1$, and so the following results hold:

$$\mathbf{J}_c \approx \frac{I(u)}{2\pi r_w} \mathbf{T} \tag{A8}$$

$$\sqrt{g} \approx r_w |\mathbf{r}'_0| \tag{A9}$$

APPENDIX B

In this Appendix we present some relevant properties of $K_{p0}(u|u')$, the periodic thin wire periodic kernel defined by (16), with $\mathbf{r}(u,\phi)$ given by (A1). The approximation (11) is used, and to keep the notation short we define $\hat{\Phi}_{p0}(\mathbf{r}|\mathbf{r}') = \Phi_{p0}(\mathbf{r}|\mathbf{r}') e^{+j\mathbf{k}\cdot(\mathbf{r}-\mathbf{r}')}$.

*Property 1*:

$$K_{p0}(u|u') = K_{p0}(u'|u) \tag{B1}$$

This property is a consequence of $\hat{\Phi}_{p0}(\mathbf{r}|\mathbf{r}') = \hat{\Phi}_{p0}(\mathbf{r}'|\mathbf{r})$.

*Property 2*:

Let us suppose that the wire parameterisation satisfies $\mathbf{r}_0(u+T) = \mathbf{r}_0(u) + \mathbf{a}$ for arbitrary u, where $\mathbf{a}$ is a primitive vector of the lattice (eventually zero) and T is a constant. Then,

$$K_{p0}(u|u') = K_{p0}(u+T|u') = K_{p0}(u|u'+T) \tag{B2}$$

In fact, $\mathbf{r}_0(u+T) = \mathbf{r}_0(u) + \mathbf{a}$ implies that $\mathbf{r}(u+T,\phi) = \mathbf{r}(u,\phi) + \mathbf{a}$, and since $\hat{\Phi}_{p0}(\mathbf{r}) = \hat{\Phi}_{p0}(\mathbf{r}+\mathbf{a})$ the property is obvious.

*Property 3*:

Let $R$ be a rotation by 90° in the horizontal plane: $R:(x,y,z)\to(-y,x,z)$. Suppose that the wire parameterisation satisfies $\mathbf{r}_0(u+u_0)=R.\mathbf{r}_0(u)+\mathbf{v}$ for arbitrary u, where $\mathbf{v}$ is a constant vector. We also consider that the lattice is orthogonal and that the transversal lattice is a square lattice. In that case, we have that:

$$K_{p0}(u|u')=K_{p0}(u+u_0|u'+u_0) \qquad (B3)$$

In fact, $\mathbf{r}_0(u+u_0)=R.\mathbf{r}_0(u)+\mathbf{v}$ implies that $\mathbf{r}(u+u_0,\phi)=R.\mathbf{r}(u,\phi)+\mathbf{v}$. Due to the lattice symmetry we also have that $\hat{\Phi}_{p0}(\mathbf{r})=\hat{\Phi}_{p0}(R.\mathbf{r})$ (i.e. $\hat{\Phi}_{p0}(\mathbf{r}|\mathbf{r}')=\hat{\Phi}_{p0}(R.\mathbf{r}+\mathbf{v}|R.\mathbf{r}'+\mathbf{v})$). Using the previous formulas we readily obtain the desired result.

*Property 4*:

Let $\tau$ be a reflection relatively to one of the coordinate axis (e.g. $\tau_x:(x,y,z)\to(-x,y,z)$). Let us suppose that the wire axis satisfies $\mathbf{r}_0(-u)=-\tau.\mathbf{r}_0(u)$ for arbitrary u, and that the lattice is orthogonal. In that case we have that:

$$K_{p0}(u|u')=K_{p0}(-u|-u') \qquad (B4)$$

Indeed, $\mathbf{r}_0(-u)=-\tau.\mathbf{r}_0(u)$ implies that $\mathbf{r}(-u,-\phi)=-\tau.\mathbf{r}(u,\phi)$. Since the lattice is orthogonal we also have that $\hat{\Phi}_{p0}(\mathbf{r})=\hat{\Phi}_{p0}(-\tau.\mathbf{r})$. Substituting these formulas into (16) we readily obtain (B4).

## APPENDIX C

Here we prove some useful properties of the integrals given by (17). To begin with, we prove that:

$$\int_{-\pi}^{\pi}\int_{-\pi}^{\pi}dudu'K_{p0}(u|u')\cos(mu)\sin(nu')=0 \qquad (C1)$$

Indeed, since the helix satisfies $\mathbf{r}_0(-u) = -\tau_x \cdot \mathbf{r}_0(u)$, we can apply property (B4) and after changing the integration variables we easily obtain (C1). Note that because of (B1) we can interchange the roles of "sin" and "cos" in the above expression.

On the other hand, we have that,

$$\int_{-\pi}^{\pi}\int_{-\pi}^{\pi} du du' K_{p0}(u|u') e^{jmu} e^{jnu'} = \int_{-\pi}^{\pi}\int_{-\pi}^{\pi} du du' K_{p0}\left(u+\frac{\pi}{2}\bigg|u'+\frac{\pi}{2}\right) e^{jmu} e^{jnu'}$$
$$= (-j)^{m+n} \int_{-\pi}^{\pi}\int_{-\pi}^{\pi} du du' K_{p0}(u|u') e^{jmu} e^{jnu'} \quad \text{(C2)}$$

where the first identity is a consequence of (B3) with $u_0 = \pi/2$, and the second identity is a consequence of (B2) and of the fact that the integrations are over one period $T = 2\pi$. From the previous result it is obvious that:

$$\int_{-\pi}^{\pi}\int_{-\pi}^{\pi} du du' K_{p0}(u|u') e^{jmu} e^{jnu'} = 0 \qquad \text{if } m+n \text{ is not multiple of 4} \quad \text{(C3)}$$

In particular, extracting the real part of the above expression and using the fact that $K_{p0}(u|u')$ is real, it follows that (assuming from now on that if $m+n$ is not multiple of 4):

$$\int_{-\pi}^{\pi}\int_{-\pi}^{\pi} du du' K_{p0}(u|u') \cos(mu) \cos(nu') = \int_{-\pi}^{\pi}\int_{-\pi}^{\pi} du du' K_{p0}(u|u') \sin(mu) \sin(nu') \quad \text{(C4)}$$

Putting $m = n = 1$, we find that, $C_1 = \tilde{C}_1$. On the other hand, replacing $n$ by $-n$ in (C4) and summing term by term the resulting identity with (C4), we obtain that:

$$\int_{-\pi}^{\pi}\int_{-\pi}^{\pi} du du' K_{p0}(u|u') \cos(mu) \cos(nu') = \int_{-\pi}^{\pi}\int_{-\pi}^{\pi} du du' K_{p0}(u|u') \sin(mu) \sin(nu') = 0 \quad \text{(C5)}$$

when both $m+n$ and $m-n$ are not multiples of 4. In particular, the above identity holds for $m \neq n$ and $0 \leq m, n \leq 2$.

## ACKNOWLEDGEMENT

This work was funded by Fundação para Ciência e a Tecnologia.